 \documentclass[a4paper,fleqn,usenatbib]{mnras}
\usepackage[T1]{fontenc}
\usepackage{ae,aecompl}

\usepackage{graphicx}	
\usepackage{amsmath}	
\usepackage{amssymb}	
\usepackage{ulem}
\usepackage{hyperref}

\title{A kinematic confirmation of the hidden Vela supercluster}

\author[Courtois et al.]
{H\'el\`ene M. Courtois$^{1}$\thanks{h.courtois@ipnl.in2p3.fr}, 
Ren\'ee C. Kraan-Korteweg$^{2}$,
Alexandra Dupuy$^{1}$,
\newauthor  Romain Graziani$^{1}$ and
Noam I. Libeskind,$^{1,3}$ \\
\\
$^{1}$University of Lyon, UCB Lyon 1, CNRS/IN2P3, IP2I Lyon, France\\
$^{2}$Department of Astronomy, University of Cape Town, Private Bag X3, 7701 Rondebosch, South Africa\\
$^{3}$Leibniz-Institut f\"ur Astrophysik Potsdam (AIP), An der Sternwarte 16, D-14482 Potsdam, Germany\\
\\
}

\begin{document}

\date{Accepted....... ;}

\pagerange{\pageref{firstpage}--\pageref{lastpage}} \pubyear{2019}

\maketitle

\label{firstpage}

\begin{abstract}
The universe region obscured by the Milky Way is very large and only future blind
large HI redshift, and targeted peculiar surveys on the outer borders will determine how much mass is
hidden there. Meanwhile, we apply for the first time two independent techniques to the galaxy peculiar velocity catalog $CosmicFlows-3$ in order to explore for the kinematic signature of a specific large-scale structure hidden behind this zone : the Vela supercluster at cz  $\sim 18,000$,km s$^{-1}$ . Using the gravitational velocity and density contrast fields, we find excellent agreement when comparing our results to the Vela object as traced in redshift space. The article provides the first kinematic evidence of a major mass concentration (knot of the Cosmic Web) located in the direction behind Vela constellation, pin-pointing that the Zone of Avoidance should be surveyed in detail in the future . 
\end{abstract}


\section{Introduction}\label{intro} 

A significant bulk flow residual was revealed in 2014 in the analysis of the 6dF peculiar velocity survey based on 8,885 galaxies in the southern hemisphere within a volume cz $\leq 16,000$\,km s$^{-1}$\  \cite{2014MNRAS.445.2677S}. The positive radial peculiar velocities suggest substantial mass over-densities originating at distances beyond that volume, three times further away than the Great Attractor at  ($ gl, gb, cz $) = (307$^\circ$, 9$^\circ$, $\sim 5000$\,km s$^{-1}$), see e.g. \cite{2016MNRAS.456.1886S} and \cite{2016AJ....151...52S}. The 6dF bulk-flow apex points to the direction of the Shapley Concentration, and close to or in the Zone of Avoidance (ZoA) in the direction of the Vela constellation.

Around the same time, redshift surveys had been launched towards the ZoA in Vela with the 10\,m Southern African Large Telescope (SALT) and AAOmega$+$2dF on the 4\,m Australian Telescope  \cite{2015salt.confE..40K,2017MNRAS.466L..29K} of optically (e.g \cite{2000A&AS..141..123K}) and near-infrared (2MASS, \cite{2000AJ....119.2498J}) identified galaxies. Earlier indications hinted at a possible excess of galaxies just beyond the 6dF volume. 
They reported the discovery of an extended supercluster of galaxies at $cz\sim 18,000$~km s$^{-1}$, straddling the Galactic Plane around $gl,gb \sim 272^\circ,0^\circ$ ($sgl,sgb \sim 173^\circ,-47^\circ$) (\cite{2017MNRAS.466L..29K,2015salt.confE..40K}, henceforth KK17a,b). 

Radial peculiar velocities can be modeled using the Wiener filter methodology  \cite{1999ApJ...520..413Z,2009LNP...665..565H,2012ApJ...744...43C}, recently re-vamped by data forward modeling and MCMC exploration \cite{2019MNRAS.tmp..130G}. This computation allows to obtain a full three-dimensional velocity field. The method has proved very successful in charting the frontiers of the Laniakea supercluster \cite{2014Natur.513...71T}, the local filament and associated planes of satellites \cite{2015MNRAS.452.1052L}, in locating the CMB cold spot repeller  \cite{Courtois2017}, and in the mapping the universe on quasi-linear scales \cite{2018NatAs...2..680H}.

In this paper, we apply the above method for an analysis of the Vela region to reveal the shape and content of the gravitational basin, and compare the results with the Vela ZoA redshift data. The independent kinematic analysis, extracted from peculiar velocity data,  is of particular importance because the core of the Vela supercluster (VSCL) remains largely uncharted at higher extinction levels.

In Sect.~2, we explore the reconstructed CF3 velocity and density field at the location of the alleged VSCL. We first examine the reconstructed fields (Sect.~2.1), and compare it to the redshift maps (Sect.~2.2), followed by a study of the streamline density (Sect.~2.3), and the velocity shear tensor (Sect.~2.4), and end with conclusions in Sect.~3.

\section{Analysis of the CF3 density contrast field in Vela}

\subsection{The reconstructed velocity and density field}

\begin{figure*}[t!]
\begin{tabular}{cc}
\includegraphics[width=0.5\textwidth]{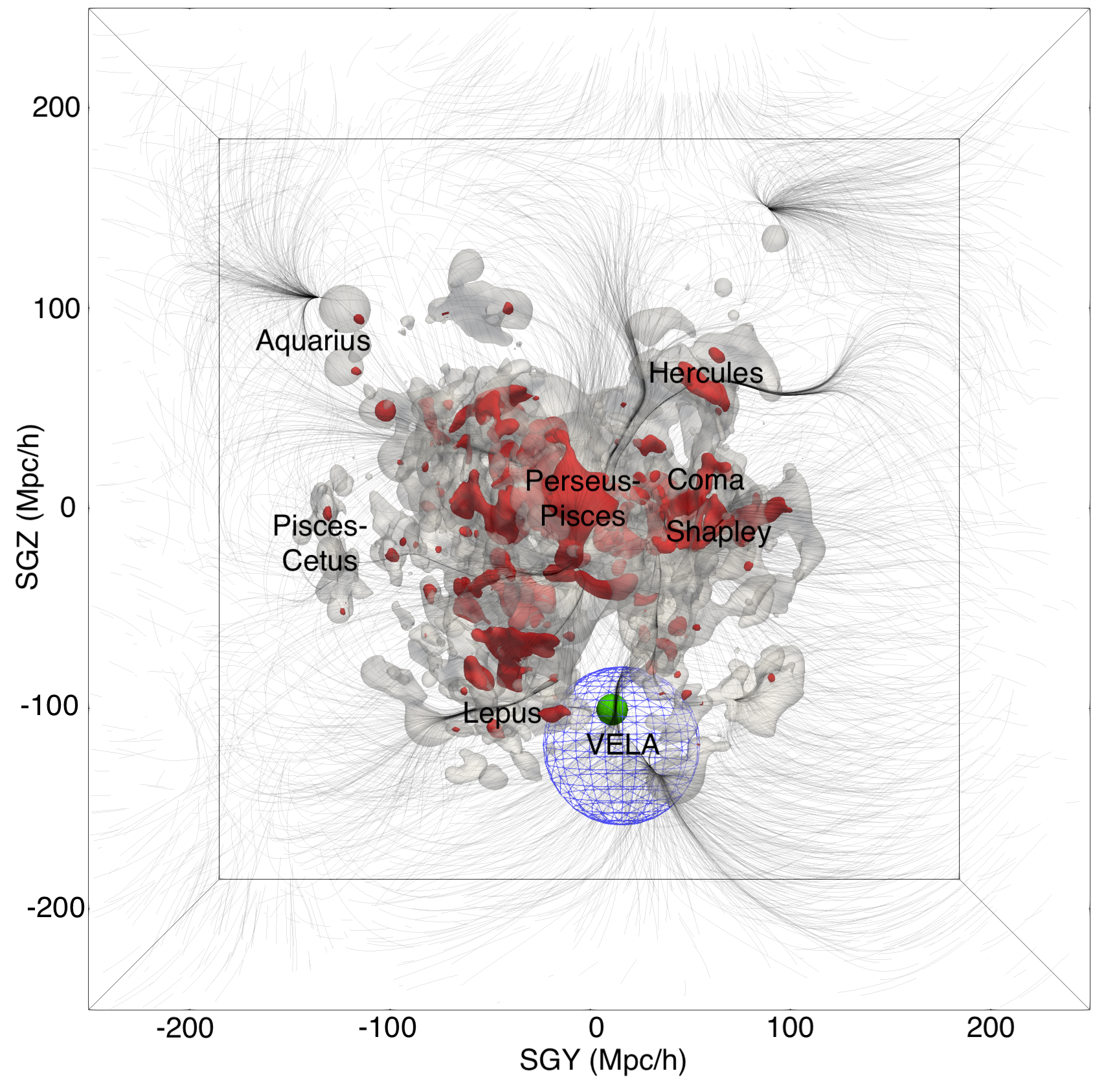}
\includegraphics[width=0.55\textwidth]{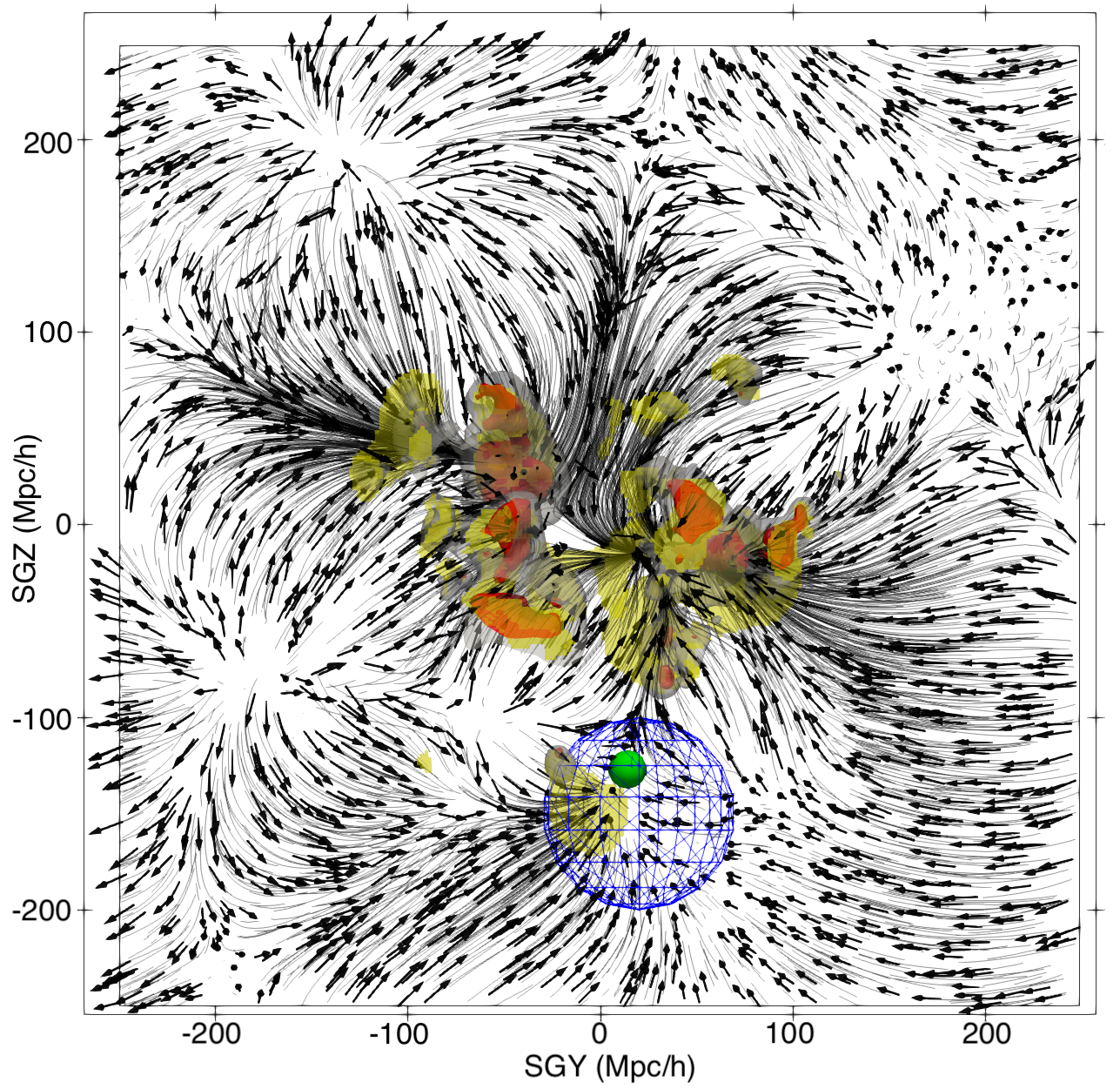}\\
\end{tabular}
\caption{ Iso-overdensity contours (left) and associated velocity field (right) in supergalactic cartesian coordinates (Mpc/$h$), as determined from CF3 \protect\cite{2019MNRAS.tmp..130G}. Left panel: the grey-shaded regions outline overdensities, the red ones the density peaks; the green dot marks the approximate VSCL position (KK17a); the blue wired sphere, centered at (SGX,SGY,SGZ)=$(-130,20,-150)$\,Mpc/$h$ of radius $R=50$\,Mpc/$h$, encompasses the Vela volume focused on in this analysis. The CF3 streamlines reveal a strong flow across the blue wired sphere. Left panel: the corresponding velocity field with the regions of compression as determined by the eigenvalues of the velocity shear tensor (see Sect. 2.3 and Fig. \ref{V_web}). 
These figures are extracted from a 3D animation which can be viewed at \url{https://vimeo.com/328453184/d3f2582f59}.} 
\label{fig-delta}
\end{figure*}

The $Cosmicflows-3$  catalog  \cite{2016AJ....152...50T} is currently the largest compilation of accurate galaxy distances independent of redshift. Compared to its precursor CF2 ($Cosmicflows-2$), CF3 includes galaxy distances from the 6dF peculiar velocity survey (6dFRSv,\cite{2012MNRAS.427..245M,2014MNRAS.443.1231C, 2014MNRAS.445.2677S}), and therefor now has excellent coverage across the major part  of  the  southern celestial hemisphere, and out to to higher distances, where CF2 had few measurements. As  such  it  is  particularly  suitable for a study of Vela overdensity: 
In the volume surrounding Vela ($225\degr \leq gl \leq 315\degr, -45\degr \leq gb \leq +45\degr$), the CF3 compendium contains 1,172 galaxies within $12,000 < V_{hel} < 15,000$~km s$^{-1}$, compared to the 96 galaxies in CF2. It is still quite sparse at the VSCL distance range (16 -- 20'000\,km s$^{-1}$; 38 in CF3 versus 10 in CF2). But note, that due to their larger coherence lengths, the velocity field (and associated flow fields such as the velocity-shear or tidal-shear fields) are more accurately chartered by such methods as the Wiener Filter than the density field. As shown in the following sections, the density of data in the CF3 compendium is sufficient to allow a reliable estimation of flow fields generated by any overdensity at the VSCL distance. This would not have been possible with VC2. 

We will now examine the CF3 reconstructed density and velocity field in the Vela region. Figure~\ref{fig-delta} displays the contours of the linear overdensity field (left panel) and corresponding velocity streamlines (right panel) computed by an iterative Monte Carlo Markov Chain exploration of parameters to allow the forward modeling of the CF3 dataset (see \cite{2019MNRAS.tmp..130G} for further details). The figure presents a slice from a movie displaying these structures in a 3-dimensional cube. The full animation can be viewed following the link given in the caption of Fig~\ref{fig-delta}.

The density contrast recovered from the peculiar velocity data set of CF3 is low at the location of VSCL, partly due to the sparse sampling at these distances, which is exacerbated by the ZoA data gap. It should be emphasised thought, that if there were no signal in the data, the model would allocate a mean density and a zero velocity field in accordance with the prior (i.e. the $\Lambda$CDM model of structure formation). The robustness and accuracy of the CF3 reconstruction are discussed thoroughly in \cite{2019MNRAS.tmp..130G}. 

While the Vela may not be very prominent in the overdensity maps, it is quite distinct in the flow field. As a matter of fact, the new reconstruction methodology introduced by \cite{2019MNRAS.tmp..130G} does not deduce the velocity field from the overdensity field as it was done by the direct Wiener Filter method. Both fields are fitted and computed individually during the convergence MCMC process. The black flow-lines in the right panel of Fig.~\ref{fig-delta} correspond to the spatially integrated linear velocity field (streamlines or flow-lines). Since the velocity field is three-dimensional, streamlines are oriented. As evident from the right panel in Fig.~\ref{fig-delta}, they flow in the direction from negative SGZ and zero SGY, to values of zero SGZ and positive SGY. The prominent convergence of flow-lines at (SGX,SGY,SGZ)=($-130,20,-150$)\,Mpc/$h$ (blue sphere in Fig.~\ref{fig-delta}; the green solid sphere marks the center as given in KK17a) is quite striking: the flow-lines run exactly through the Vela complex as described by the redshift survey, which constitutes a completely independent data set. This is one of the salient points of this study: {\sl the velocity field reconstructed from the CF3 catalog of peculiar velocities points to an attractor independently predicted by redshift surveys}.

\subsection{Matching reconstructions to redshift maps}
Based on the encouraging results, we will compare in our reconstructions derived from the peculiar velocity data set with the available redshift data in this region (both from KK17a and the literature) in further detail.

\begin{figure*}
\includegraphics[width=1\textwidth]{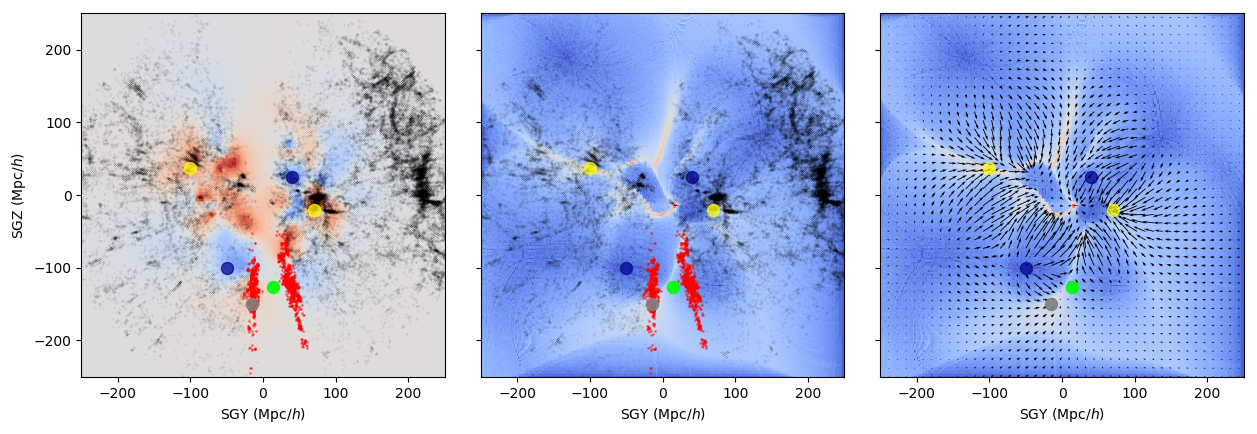}
\caption{Comparison of the $\delta$ density contrast field (left) and the image of streamlines density (middle and right) based on CF3 The figures are in the supergalactic SG-YZ plane, centered at SGX$=-130$\,Mpc/h (Vela position). Red dots represent galaxies from the dedicated Vela redshift survey, black dots from the literature. The green sphere marks the VSCL centre according to KK17a. VSCL has a weak signal in the density maps due to the scarcity of observational data in the ZoA, whereas the kinematics reveal a strong coherent flow across the ZoA  at the Vela position. See animated figures of the delta field at \url{https://vimeo.com/327932938/cde7531fc5}, and of the streamline density at \url{https://vimeo.com/327932920/2912c2d1fe}. The large yellow, blue and grey dots highlight kinematically identified under-, respectively overdensities in this projection, discussed further in Sect.~2.3.}
\label{fig_slices_vela}
\end{figure*}

The redshift surveys conducted in the southern hemisphere in the wider Vela region (KK17a,b)  were designed to explore galaxy structures as close as possible to ZoA, i.e. up to extinction and star density levels that precludes optical spectroscopy. These redshifts are displayed as red dots in Fig.~2 (left and middle panels) together with redshifts compiled in the literature (black dots). Despite the sparse redshift sampling to date (see Fig.~1 in KK17a) -- and the remaining ZoA redshift data gap (visible along the SGY=0-axis) -- a clear over-density of galaxies is seen at SGZ$=-130$~Mpc/$h$ on both sides of the ZoA. This is the signature of what was identified as the main Vela SCL Wall (Wall~1), a structure spanning over 60\,Mpc/$h$ across the ZoA. The VSCL redshift survey also revealed a second wall (Wall~2) visible at slightly higher redshifts below the Galactic Plane. It is suspected to merge with the lower redshift Wall in the opaque part of the Milky Way (see Fig.~3 in KK17a).
Based on the location of the two walls identified in the Vela ZoA redshift survey, in combination with spectroscopic and photometric redshift data adjacent to the Vela ZoA region (e.g. Fig.~13 in the 2MPZ; \cite{2014ApJS..210....9B}), the likely centre of this mass overdensity was predicted to lie around $(gl,gb,cz) \sim (272\degr,0\degr,18,000$\,km s$^{-1}$), corresponding  to  (SGX,SGY,SGZ)$\sim(-126,14,-127)$\,Mpc/$h$ in supergalactic cartesian coordinates (green circle in Fig.~2), hence in between the two largest galaxy concentrations on either side of the Galactic Plane. This location also coincides with the suspected location of the crossing of the two Vela walls; it is not inconceivable that the ZoA hides a dense central core in its fully obscured part, as so often is seen at the junction of walls. 

The kinematic CF3-analysis finds the centre of the Vela overdensity {\sl only} 10~degrees away from this position, at a marginally higher distance, i.e. at $(gl, gb, cz)$=$(269\degr,-9\degr,19,900$\,km s$^{-1}$) or (SGX,SGY,SGZ)=($-130,-15,-150)$\,Mpc/$h$; marked as a grey dot in Fig.~\ref{fig_slices_vela}. The accuracy on superclusters positions in CF3 is not better than 3 Mpc/$h$ due to the grid resolution, and the variance in position is at the same level between different constrained realizations. This proximity between the determination of the central basin in both methods is remarkable given the different intrinsic uncertainties in both methods. This slight offset could well be the influence of the second Vela wall, which has a mean redshift of $cz\sim21,000$,km s$^{-1}$, and is more dominant below the Galactic Plane, but weak at positive Galactic latitudes. The position of the centre of the potential well of Vela as found in the CF3 analysis seems to support the interpretation in KK17a which suggests the Vela overdensity to be a supercluster in formation, with two merging wall-like structures.

In addition to the positions of galaxies in redshift space, the left hand panel of Fig.~\ref{fig_slices_vela} also displays the CF3 density contrast field (the $\delta-field$) illustrating the high- and low-density areas with different colours. We used the density field to identify areas of high-densities derived with our analysis in this plane. Apart from Vela (green dots) two other overdensities (yellow dots) and underdensities (blue dots) stand out. They will be discussed further in the next section. In the obscuration zone, not much signal is recovered in terms of the $\delta-field$, due to the lack of observational data in this part of the  sky.

\subsection{Analysis of the CF3 streamline density}
We now move to a kinematic study of the velocity field. This includes a study of the streamlines, followed in the next section with an analysis of the velocity shear. 
As shown by \cite{Dupuy2019}, the velocity vectors can be followed across the grid to build up streamlines; these allow a segmentation of the local universe into gravitational basins according to the ubiquity of stream lines. To determine these statistics, we use a 256$^3$ grid placed on a 320\,Mpc/$h$ box centered at the observer. The method uses a present day static universe to compute the flow lines from the discretized, reconstructed peculiar-velocity field. 

The colour-coded intensity of such streamlines -- derived from the sums in the grid are displayed in the middle and right-hand panels of Fig.~\ref{fig_slices_vela}. The white streams illustrate the areas of highest concentrations of streamlines while the blue regions have fewer streamlines, and as such actually mark the main flowing patterns of cosmic matter. This is similar to representations using velocity field arrows, as evident in the right most panel of Fig.2 which displays both. When comparing the high-density streamlines with the known galaxy redshift locations, it is immediately apparent that the predicted location of the Vela filament (according KK17a), coincides with a high concentration of streamlines. The streamline method provides further evidence for the partially hidden Vela mass overdensity.

Nevertheless, the signal in the velocity field is relatively weak  because of the large distance to VSCL. To get a better understanding of the robustness of this result, we investigate four other regions located in the same SGX-plane, which are at the approximate same distance. We identified two overdense and two underdense regions by inspecting the known galaxy distribution (redshift surveys) and the density contrast field. They are marked in Fig.~2 as yellow and blue dots respectively. Table \ref{tab_values_spheres} lists the values of the density contrast ($\delta$) and the number of streamlines crossing the grid cell at the four test locations. The underdense regions, unsurprisingly, are very low in terms of the density contrast ($\delta\sim-0.9$) and the number of streamlines. The dense regions have $\delta\sim1$ with a high number of streamlines. It is interesting to find these overdensities, in hindsight, listed in the supercluster catalog of \cite{2014MNRAS.445.4073C} as SSCC~292 and SSCC~249, with $z\sim0.052$ and $0.054$ respectively. Note that SSCC~249 lies very close to the Shapley core; its very high density and streamline values might be influenced by this proximity. 

The density peak and streamline cell counts of VSCL are lower ($\delta\approx 0.4$ $\pm 0.2$ and N=449). Errors bars have been estimated at the Vela position (within 60 Mpc/$h$) using 100 CF3 constrained realizations. This is also true for Shapley. However, the numbers obtained are well above the mean density and support the existence of a large mass excess hidden behind the ZoA. Table~1 also gives the extent of these structures at which their overdensity drops to half its value, which shows that both Vela and Shapley are extended in comparison to the other superclusters, as well as the Great Attactor.

\begin{table*}
\caption{Properties of three significant superclusters in the local universe (Vela, Great Attractor and Shapley), with two high-density and two void regions at similar distance as Vela (see Fig.~2). The properties are drawn from an analysis of the CF3 reconstructed peculiar velocity and density field. We give cartesian supergalactic coordinates, $\delta$ values at each the over and underdensities, the distance from this peak to half of this peak-density, and to the zero contrast (mean universe density), and the number of streamlines at the peak (computed according to \protect\cite{Dupuy2019}).
Distances given in columns 6 and 7 corresponds to the radius of a sphere centered on the peak density density.}
\begin{center}
\begin{tabular}{l r r r @{\hskip 0.4in} l l l @{\hskip 0.2in} l }
 Structure      & SGX  & SGY  & SGZ &  $\delta$ & dist. to & dist. to  & \# streamlines     \\
                &Mpc/$h$ &Mpc/$h$ &Mpc/$h$    &  peak & $\delta=0$ & $\delta=\delta_{peak}/2$ &  peak    \\
       &   &   &  &   & Mpc/$h$  & Mpc/$h$  &      \\

\hline
Vela            & -130 &  -15 & -150 &  0.4 & 60 & 40 &  449    \\
Great Attractor &  -49 &    4 &    8 & 2.0  & 30 & 18 & 2.4e7   \\
Shapley         & -146 &   53 &  -23 & 1.6  & 35 & 20 & 9.1e6     \\
Void 1          & -130 &  40  &   25 & -1.5 & 30 & 10 &     7   \\
Void 2          & -130 & -50  & -100 & -0.9 & 70 & 40 &     3   \\
high density 1  & -130 & -100 &   37 &  0.9 & 25 & 10 &   700   \\
high density 2  & -130 &   70 &  -20 &  1.7 & 30 & 7  & 12400  \\
\end{tabular}
\end{center}
\label{tab_values_spheres}
\end{table*}%

\subsection{Analysis of the CF3 velocity shear tensor}

\begin{figure*}
\includegraphics[width=1\textwidth]{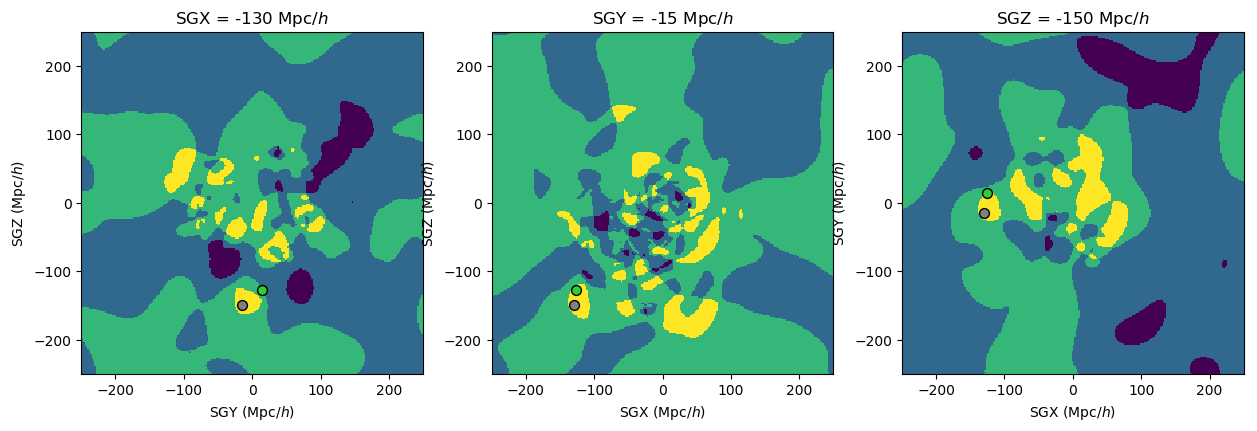}
\caption{From left to right we show the three SGY-SGZ, SGX-SGZ, and SGX-SGY slices at the putative location of VSCL (see Table~1). Knots, filaments, sheets and voids are shown in yellow, green, light blue,  and dark blue respectively. They are derived from the velocity field by counting the number of positive eigenvalues of the diagonalized velocity shear tensor. At the position of Vela a region of compression along all three axes is found, implying accretion onto the overdensity.}
\label{V_web}
\end{figure*}

In this section we analyse the cosmic velocity web (following \cite{2012MNRAS.425.2049H,2014MNRAS.441.1974L}) derived from the CF3 peculiar velocity field. The velocity shear tensor is computed as $\Sigma_{\alpha\beta}=-\frac{1}{2H_{0}}\big(\frac{\partial v_{\alpha}}{\partial r_{\beta}}+\frac{\partial v_{\beta}}{\partial r_{\alpha}}\big)$, where $\alpha$ and $\beta$ are the x, y, and z components. The $H_{0}$ normalization makes the quantity dimensionless and the minus sign is used to make the positive eigenvectors correspond to a converging flow. The shear tensor is diagonalised at each grid point to obtain the eigenvalues $\lambda_{i}$ and the corresponding eigenvectors $\hat{e}_{i}$, (where $i=1,2,3$ and $\lambda_{1}>\lambda_{2}>\lambda_{3}$). The web classification at each grid cell is done by counting how many eigenvalues are positive thus segmenting the velocity shear field into knots (3 positive eigenvalues), filaments (2 positive eigenvalues), sheets (1 positive eigenvalue) and voids (all negative eigenvalues). Figure~\ref{V_web} shows the knots (in yellow), filaments (in light green), sheets (in purple) and voids (in dark green). Note that all knots are connected by filaments. At the position of VSCL predicted by the redshift surveys, a well-defined  knot (yellow clump) is found embedded in a filament (light green).  This is a confirmation that the volume encompassing Vela is experiencing a compression along all three directions. Compression along three orthogonal directions is a clear signature of accretion, supporting the hypothesis of VSCL being a supercluster in formation.

Defining the extent of the VSCL by examining the knot region in which it is embedded yields a mean density of 0.34 and a size of $\sim60$\,Mpc/$h$ across. This is in very close agreement with the streamline convergence discussed above. 

\section{Conclusion}

After redshift surveys alleged the existence of a vast structure hidden behind the ZoA in the direction of Vela, dubbed the Vela supercluster, an analysis based on peculiar velocities in this general region became an exciting option for confirming this structure. An analysis comparing redshift and peculiar velocity surveys had never been done before for an extended large-scale structure at such high distances. The CF3 compendium opened up this opportunity for the very first time.

The derivations from the observed peculiar-velocity field presented in this paper provide clear confirmation of Vela structure. The velocity field indicates the mass concentration to be located at $(gl,gb,cz)$=$269\degr,-9\degr,19,900$\,km s$^{-1}$, which is in close proximity to the results presented in KK17a. The small positional and redshift offset between the two estimates ($\sim10\degr$ and redshift $\sim2000$\,km s$^{-1}$) could well be induced by the merging second higher-redshift wall that forms part of the supercluster.

The kinematic analysis suggest the VSCL to encompass a volume of the order of 60\,Mpc/$h$ with an overdensity around $\sim 0.4$. This is  closely aligned with the overdensity derived in KK17a for the VSCL volume based on photometric redshifts of 2MASS galaxies along the less obscured part of the ZoA ($|gb| =6\degr-10\degr$), and was found to be similar to the Shapley overdensity when subjected to same biases, which corresponds to the findings here. 

The fact that the existence of the Vela overdensity is confirmed through its gravitational effect derived by the peculiar velocity field from unobscured data makes its discovery all the more meaningful, particularly when considering that a major fraction of VSCL remains hidden in opaque part of the ZoA. This kinematic method demonstrates its promise for further discoveries. In more general terms, the universe is being continously mapped further and consequently the volume of Universe hidden by the ZOA is becoming larger and larger, encompassing tens of hidden major cosmic web features. Just like the unveiling of the Vela supercluster more features can be discovered by the presented  methodology. This article is calling for suuport to MeerKAT HI redshifts observations of the inner opaque ZOA, complemented with a systematic survey of peculiar velocity observations with WALLABY and TAIPAN along the outer borders of the ZoA.

\section*{Acknowledgments}

Support has been provided by the Institut Universitaire de France, the CNES, and the Israel Science Foundation (1013/12). NIL and AD acknowledge financial support of the Project IDEXLYON at the University of Lyon under the Investments for the Future Program (ANR-16-IDEX-0005). RKK thanks the South African National Research Foundation for their support. We are grateful to the colleagues maintaining LEDA Lyon extragalactic database Dmitry Makarov and Philippe Prugniel.

\bibliographystyle{mnras}
\bibliography{paper}

\bsp	
\label{lastpage}

\end{document}